\documentclass[
   ,final            
   ,numberedheadings 
  ]
  {aipproc}

\layoutstyle{6x9}

\begin{document}

\title{Radio Observations  of Supernovae}


\author{Nino Panagia\footnote{Also: Istituto Nazionale di Astrofisica
(INAF), Via del Parco Mellini 84, I-00136, Rome, Italy; and Supernova
Ltd., OYV \#131, Northsound Road, Virgin Gorda, British Virgin
Islands.}~}{ address={STScI, 3700 San Martin Dr., Baltimore, MD 21218,
USA; \texttt{panagia@stsci.edu}} 
}

\author{Kurt W. Weiler}{
  address={Naval Research Laboratory, Code 7210, Washington, DC
  20375-5320, USA; \texttt{Kurt.Weiler@nrl.navy.mil}}
}

\author{Schuyler D.~Van Dyk}{
  address={Spitzer Science Center/Caltech, MS 100-22, Pasadena, CA 91125, USA; \texttt{vandyk@ipac.caltech.edu}}
}
\author{Richard A. Sramek}{
  address={NRAO/VLA, PO Box 0, Socorro, NM 87801 USA; \texttt{dsramek@nrao.edu}}
  }
  
\author{Christopher J. Stockdale}{
  address={Marquette University, Physics Dept., PO Box 1881, Milwaukee,
  WI  53201-1881, USA; \texttt{Christopher.Stockdale@marquette.edu}}
  }

\begin{abstract}

Study of radio supernovae over the past 25 years includes two dozen
detected objects and more than 100 upper limits. From this work it is
possible to identify classes of radio properties, demonstrate
conformance to and deviations from existing models, estimate the density
and structure of the circumstellar material and, by inference, the
evolution of the presupernova stellar wind.  It is also possible to
detect ionized hydrogen along the line of sight, to demonstrate binary
properties of the stellar system, to detect clumpiness of the
circumstellar material, and to put useful constraints  on the
progenitors of undetected Type Ia supernovae.

\end{abstract}

\maketitle


\section{Radio Supernovae}

A series of papers on radio supernovae (RSNe) has established the radio
detection and, in a number of cases, radio evolution for approximately
two dozen type~Ib/c supernovae (SNe) (Because the differences between
the SN optical classes are slight -- type~Ib show strong He I absorption
while type~Ic show weak He I absorption -- and there are no obvious
radio differences, we shall often refer to the classes as type~Ib/c.),
and the rest type~II SNe.  A much larger list of more than 100
additional SNe have low radio upper limits (See {\it
http://rsd-www.nrl.navy.mil/7213/weiler/kwdata/rsnhead.html}).

In this extensive study of the radio emission from SNe, several
effects have been noted: 

\begin{itemize}
\item[-] {type~Ia SNe are not radio emitters to the detection limit of
the VLA\footnote{The VLA telescope of the National Radio Astronomy
Observatory is operated by Associated Universities, Inc.\ under a
cooperative agreement with the National Science Foundation.}}
\item[-] {type~Ib/c SNe are radio luminous with steep spectral indices
(generally $\alpha < -1$; $S \propto \nu^{+\alpha}$) and have a fast
turn-on/turn-off, usually peaking at 6 cm near or before optical
maximum.}
\item[-] {type~II SNe show a range of radio luminosities with flatter
spectral indices (generally $\alpha > -1$) and a relatively slow
turn-on/turn-off, usually peaking at 6 cm significantly after optical
maximum.}
\end{itemize}

Measurements of the multi-frequency radio light curves and their
evolution with time show the density and structure of the CSM,
evidence for possible binary companions, clumpiness or filamentation
in the presupernova wind, mass-loss rates and changes therein for the
presupernova stellar system and, through stellar evolution models,
estimates of the ZAMS presupernova stellar mass and the stages through
which the star passed on its way to explosion.

\section{Emission Models}

All known RSNe appear to share common properties of: 

\begin{itemize}
\item[-] {Nonthermal synchrotron emission with high brightness temperature.}
\item[-] {A decrease in absorption with time, resulting in a smooth,
rapid turn-on first at shorter wavelengths and later at longer
wavelengths.}
\item[-] {A power-law decline of the flux density with time at each
wavelength after maximum flux density (optical depth $\sim 1$) is
reached at that wavelength.}
\end{itemize}

The characteristic RSN radio light curves arise from the competing
effects of slowly declining non-thermal radio emission and more rapidly
declining thermal or non-thermal absorption yielding a rapid turn-on and
slower turn-off of the radio emission at any single frequency.  This
characteristic light curve shape is illustrated in Fig.~1-2 for
SN 1979C and SN 1980K. Since absorption processes are greater at lower
frequencies, transition from optically thick to optically thin (turn-on)
occurs first at higher frequencies and later at lower frequencies.

\begin{figure}
  \includegraphics[height=.35\textheight,angle=-90]{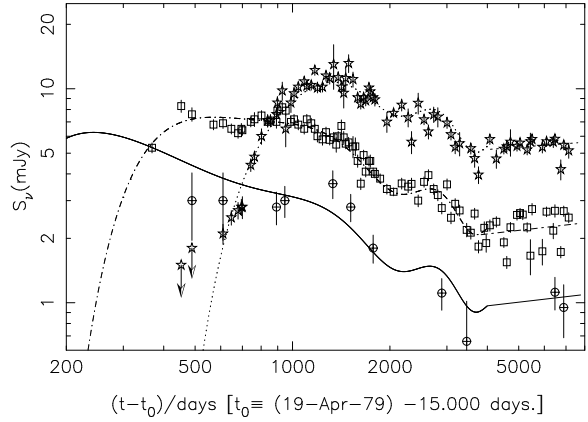}
  \caption{Type II SN~1979C at 2 cm (14.9 GHz; {\it crossed circles, solid line}), 
6 cm (4.9 GHz; {\it open squares, dash-dot line}), and 20 cm (1.5 GHz;
{\it open stars, dotted line}).  Note that the radio flux density
increases after day $\sim4000$ and has a sinusoidal modulation before
day  $\sim4000$ (Weiler {\it et al.} 
1991, 1992a, and Montes {\it et al.} 
2000). }
\end{figure}

\begin{figure}
  \includegraphics[height=.35\textheight,angle=-90]{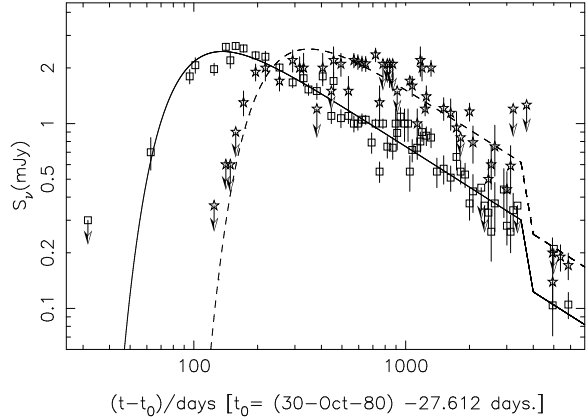}
  \caption{Type II SN~1980K at 6 cm
(4.9 GHz; {\it open squares, solid line}), and 20 cm (1.5 GHz; {\it open
stars, dashed line}). Note a sharp drop in flux density after day $\sim
4000$ (Weiler {\it et al.} 1992b, and Montes {\it et al.}
1998).}
\end{figure}

Chevalier (1982a, 1982b) has proposed that the relativistic electrons
and enhanced magnetic field necessary for synchrotron emission arise
from the SN blastwave interacting with a relatively high density CSM
which has been ionized and heated by the initial UV/X-ray flash.  This
CSM density ($\rho$), which decreases as an inverse power, s, of the
radius, r, from the star, is presumed to have been established by a
presupernova stellar wind with mass-loss rate, $\dot M$, and velocity,
$w_{\rm wind}$, (i.e., $\rho \propto \frac{\dot M}{w_{\rm wind}~r^2}$)
from a massive stellar progenitor or companion.  For a constant
mass-loss rate and constant wind velocity $\rho \propto r^{-2}$.  This
ionized CSM is the source of some or all of the initial thermal gas
absorption.  Additionally, Chevalier (1998) has proposed that
synchrotron self-absorption (SSA) may play a role in some objects.  

A rapid rise in the observed radio flux density results from a decrease
in these absorption processes as the radio emitting region expands and
the absorption processes, either internal or along the line-of-sight,
decrease.  Weiler {\it et al.} (1990) have suggested that this
CSM can be ``clumpy'' or ``filamentary,'' leading to a slower radio
turn-on, and Montes {\it et al.} (1997) have proposed the
possible presence of a distant ionized medium along the line-of-sight
that is sufficiently distant from the explosion that it is unaffected by
the blastwave and can cause a spectral turn-over at low radio
frequencies.  In addition to clumps or filaments, the CSM may be
radially structured with significant density irregularities such as
rings, disks, shells, or gradients.

\section{Parameterized Radio Light Curves}

Weiler {\it et al.} (1986, 1990) and Montes {\it et al.} (1997) adopted 
a parameterized model which has been updated in Weiler {\it et al.}
(2002).  The model includes an intrinsic sychrotron emission, which is
characterized by a spectral index $\alpha<0$ and declines with time as a
power law with exponent $\beta<0$ ($S_\nu \propto \nu^\alpha t^\beta$),
and an attenuation mostly produced by {\it f-f} absorption from the ionized
CSM, but also, especially at early times, by sychrotron 
self-absorption.  Each {\it i-th} absorption process is characterized by
a power law decline with  slope $\delta _i$. Various possibilities for
the  {\it f-f} absorption are included, namely (i) homogeneous and (ii)
clumpy or filamentary absorption by gas in the CSM, as well as (iii) the
one originating in  a possible foreground HII region. The homogeneous 
absorption is produced by an ionized medium that completely covers the
emitting source, and additional attenuation  may be produced by an
inhomogeneous medium (``clumpy absorption''; see Natta \& Panagia
1984 for a more detailed discussion of attenuation in
inhomogeneous media). The distant {\it f-f} absorption is produced by a
homogeneous medium which completely covers the source but is so far from
the SN progenitor that it is not affected by the expanding SN blastwave
and is constant in time.  

Since it is physically realistic and may be needed in some RSNe where
radio observations have been obtained at early times and high
frequencies, the model  also includes the possibility for an internal
absorption term.  This internal absorption  term consists  of two parts
-- synchrotron self-absorption, and mixed, {\it f-f} absorption due to
thermal gas coexisting with the  non-thermal emitting electrons.

\section{Results } 

The success of the basic parameterization and modeling has been shown in
the good correspondence between the model fits and the data for all
subtypes of RSNe: e.g., type Ib SN~1983N (Sramek {\it et al.} 1984),
type Ic SN~1990B (Van Dyk {\it et al.} 1993a), and type II SN~1979C 
(Weiler {\it et al.} 1991, 1992a, and Montes {\it et al.} 2000) and
SN~1980K (Weiler {\it et al.} 1992b, and Montes {\it et al.} 1998). 
Note that after day $\sim4000$, the evolution of the radio emission from
both SN~1979C and SN~1980K deviates from the expected model evolution
and that SN~1979C shows a sinusoidal modulation in its flux density
prior to day $\sim$4000 (se Fig. 1 and 2).

Additionally, in their study of the radio emission from SN~1986J, Weiler
{\it et al.} (1990)  found that the simple Chevalier model could not
describe the relatively slow turn-on.  They therefore included terms
described mathematically by a clump optical depth,  $\tau_{{\rm
CSM}_{\rm clumps}}$, in the model equations.  This extension greatly
improved the quality of the fit and was interpreted by Weiler {\it et
al.} (1990) to represent the presence of filaments or clumps in the
CSM.  Such a clumpiness in the wind material was again required for
modeling the radio data from SN~1988Z (Van Dyk {\it et al.} 1993b, and
Williams {\it et al.} 2002) and SN~1993J (Van Dyk {\it et al.} 1994). 
Since that time, evidence for filamentation in the envelopes of SNe has
also been found from optical and UV observations (Filippenko {\it et
al.} 1994, and Spyromilio 1994).    

\subsection{Mass-Loss Rate from Radio Absorption}

From the Chevalier model (1982a, 1982b), the turn-on
of the radio emission for RSNe provides a measure of the presupernova
mass-loss rate to wind velocity ratio ($\dot M/w_{\rm wind}$).  Weiler
{\it et al.} (1986) derived this ratio for the case of pure,
external absorption by a homogeneous medium.  However, Weiler {\it et
al.} (2002) propose several possible origins for absorption and
generalize Eq.~16 of  Weiler {\it et al.} (1986) to

\begin{eqnarray}
\label{eq1}
\frac{\dot M ({\rm M_\odot} ~ {\rm yr}^{-1})}{( w_{\rm wind} / 10\ {\rm km\ s}^{-1} )} = 3.0 \times 10^{-6}\ <\tau_{{\rm eff}}^{0.5}>    m^{-1.5} {\left(\frac{v_{\rm i}}{10^{4}\ {\rm km\ s}^{-1}}\right)}^{1.5} {\left(\frac{t_{\rm i}}{45\ {\rm days}}\right) }^{1.5} \nonumber \\ \times 
 {\left(\frac{t}{t_{\rm i}}\right) }^{1.5 m}  {\left(\frac{T}{10^{4}\ {\rm K}} \right)}^{0.68}
\end{eqnarray}

\noindent where $m$ is the exponent of the power-law growth of the SN
front radius, $r\propto t^m$.

Since the appearance of optical lines for measuring SN ejecta velocities
is often delayed a bit relative to the time of the explosion, for
convenience they take the reference time to be $t_{\rm i}$ = 45 days. 
Also,  because many SN measurements indicate velocities of $\sim10,000$
km s$^{-1}$, one usually assumes $v_{\rm i} = v_{\rm blastwave} =
10,000$ km s$^{-1}$\ and takes values of $T = 20,000$ K, $w_{\rm wind} =
10$ km s$^{-1}$\ (which is appropriate for a RSG wind), $t = (t_{\rm
6cm\ peak} - t_0)$ days from best fits to the radio data for each RSN,
and $m = -\delta/3$ or $m = -(\alpha - \beta - 3)/3$, as appropriate
(Weiler {\it et al.} 2002, and Sramek \& Weiler 2003).  

The optical depth term $<\tau_{{\rm eff}}^{0.5}>$ used by Weiler {\it et
al.} (1986) is extended by Weiler {\it et al.} (2001)
and they identify at least three possible absorption regimes: 1)
absorption by a homogeneous external medium, 2) absorption by a clumpy
or filamentary external medium with a statistically large number of
clumps, and 3) absorption by a clumpy or filamentary medium with a
statistically small number of clumps.  These three cases have different
formulations for $<\tau_{{\rm eff}}^{0.5}>$ which are described in
detail by Weiler {\it et al.} (2002). Mass-loss rate
estimates from radio absorption obtained in this manner tend to be
$\sim10^{-6}\ {\rm M_\odot} ~ {\rm yr}^{-1}$ for type Ib/c SNe and
$\sim10^{-4} - 10^{-5}\ {\rm M_\odot} ~ {\rm yr}^{-1}$ for type II SNe. 
Estimates for some of the best studied SNe are given in Weiler {\it et
al.} (2002).  As discussed by Panagia {\it et al.} (2006, in
preparation), the high mass loss rates measured for SNII confirm that
their progenitors are Red Supergiants with original masses above 8 
${\rm M_\odot}$.

\subsection{Changes in Mass-Loss Rate\label{changes}}

A particularly interesting case of mass-loss from an RSN is SN~1993J,
where detailed radio observations are available starting only a few days
after explosion (Fig.~3).  Van Dyk {\it et al.}
(1994) find evidence for a changing mass-loss rate
(Fig.~4) for the presupernova star which was as high as
$\sim10^{-4}\ {\rm M_\odot}\ {\rm yr}^{-1}$ approximately 1000 years
before explosion and decreased to $\sim10^{-5}\ {\rm M_\odot}\ {\rm
yr}^{-1}$ just before explosion, resulting in a relatively flat density
profile of $\rho \propto r^{-1.5}$.

Fransson \& Bj\"orgsson (1998) have suggested that the
observed behavior of the {\it f-f} absorption for SN~1993J could alternatively
be explained in terms of a systematic decrease of the electron
temperature in the circumstellar material as the SN expands.  It is not
clear, however, what the physical process is which determines why such a
cooling might occur efficiently in SN~1993J, but not in SNe such as
SN~1979C and SN~1980K where no such behavior is seen.  Also, recent
X-ray observations with the ROSAT of SN~1993J indicate a non-$r^{-2}$
CSM  density surrounding the SN progenitor (Immler {\it et al.}
2001), with a density gradient of $\rho \propto r^{-1.6}$ .

\vspace{1cm}
\begin{figure}[!htb]
  \includegraphics[height=0.3\textheight,angle=-90]{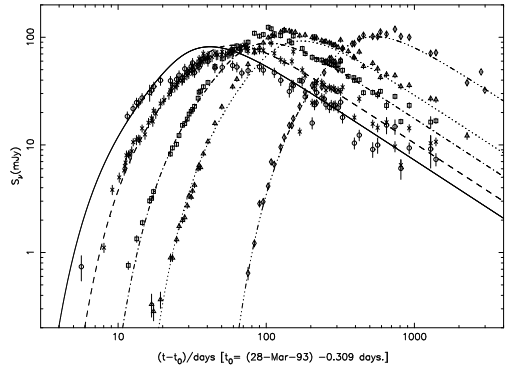}
  \caption{Type IIb SN~1993J at 1.3 cm (22.5 GHz; {\it open
circles, solid line}), 2 cm (14.9 GHz; {\it stars, dashed line}), 3.6 cm
(8.4 GHz; {\it open squares, dash-dot line}), 6 cm (4.9 GHz; {\it  open
triangles, dotted line}), and 20 cm (1.5 GHz; {\it open diamonds,
dash-triple dot line}).}
\end{figure}


\begin{figure}[!htb]
  \includegraphics[height=0.3\textheight]{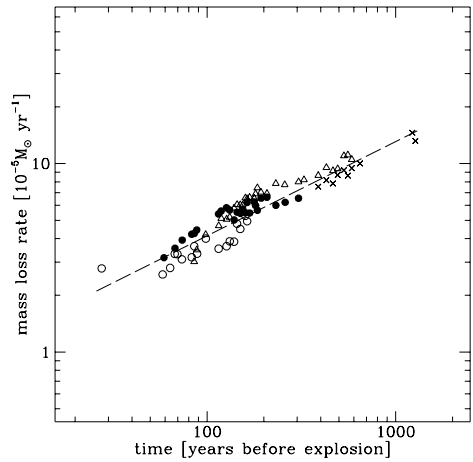}
  \caption{Changing mass-loss rate of the
presumed red supergiant progenitor to SN~1993J versus time before the
explosion.}
\end{figure}

Moreover, changes in presupernova mass-loss rates are not unusual.  
Montes {\it et al.} (2000) find that type II SN~1979C had a
slow increase in its radio light curve after day $\sim4300$ (see
Fig.~1) and type II SN~1980K showed a steep decline in flux
density at all wavelengths (see Fig.~2) by a factor of $\sim2$
occurring between day $\sim$3700 and day $\sim$4900 Montes {\it et al.}
(1998).

\subsection{Binary Systems}

In the process of analyzing a full decade of radio measurements from
SN~1979C, Weiler {\it et al.} (1990, 1992a) found evidence
for a significant, quasi-periodic, variation in the amplitude of the
radio emission at all wavelengths of $\sim 15\%$ with a period of 1575
days or $\sim 4.3$ years (see Fig.~1 at age $<$ 4000 days). 
They interpreted the variation as due to a minor ($\sim 8\%$) density
modulation, with a period of $\sim 4000$ years, on the larger,
relatively constant presupernova stellar mass-loss rate.  Since such a
long period is inconsistent with most models for stellar pulsations,
they concluded that the modulation may be produced by interaction of a
binary companion in an eccentric orbit with the stellar wind from the
presupernova star.

This concept was strengthened by more detailed calculations for a binary
model from Schwarz \& Pringle (1996).  Since that time, the presence of
binary companions has been suggested for the progenitors of SN~1987A
(Podsiadlowski {\it et al.} 1992), SN~1993J (Podsiadlowski {\it et al.}
1993), SN~1994I (Nomoto {\it et al.} 1994), SN~1998bw (Weiler {\it et
al.} 2001),  and SN~2001ig (Ryder {\it et al.} 2004), indicating that
binaries may be common in presupernova systems.

\subsection {Ionized Hydrogen Along the Line-of-Sight}

A reanalysis of the radio data for SN~1978K from  Ryder {\it et al.}
(1993) clearly shows flux density evolution characteristic of
normal type II SNe.  Additionally, the data indicate the need for a
time-independent, free-free absorption component.   Montes {\it et al.}
(1997) interpreted this constant absorption term as indicative of
the presence of HII along the line-of-sight to SN~1978K, perhaps as part
of an HII region or a distant circumstellar shell associated with the SN
progenitor. 
A high-dispersion spectrum of SN~1978K at the wavelength range
$6530-6610$ \AA\  obtained by Chu {\it et al.} (1999) showed
narrow nebular H$\alpha$ and [N II] emission with a high [N II]
6583/H$\alpha$\ ratio of $0.8-1.3$ indicative of a stellar ejecta
nebula. These data suggest that the nebula detected towards SN~1978K is
probably part of a large, dense, structured circumstellar envelope. 

\subsection {Constraints on Type Ia Supernova Progenitors}

Recently Panagia {\it et al.} (2006) have discussed the radio
observations  of 27 Type Ia supernovae (SNIa) observed over two
decades with the Very Large Array.  No SNIa has been detected so far in
the radio, implying a very low density for any possible circumstellar
material established by the progenitor, or progenitor system, before
explosion.  They derive $2\sigma$ upper limits to a steady mass-loss
rate for individual SNIa systems as low as $\sim 3 \times 10^{-8}$
M$_\odot$ yr$^{-1}$, discriminating strongly against white dwarf
accretion via a stellar wind from a massive binary companion in the
symbiotic star. However, in  the ``single degenerate'' scenario, 
a white dwarf accreting from a relatively low mass companion
via a high efficiency ($>60-80$ \%) Roche lobe overflow is
still consistent with such limits. The ``double degenerate'' merger
scenario also cannot be excluded.

\section{Conclusions}

The success of the basic parameterization and modeling is shown in the
good agreement between the model fits and the data for all subtypes of
RSNe. Thus, the radio emission from SNe appears to be relatively well
understood in terms of blastwave interaction with a structured CSM and
allows description of such progenitor system properties as mass-loss
rate, change in mass-loss rate, filamentation or clumpiness, binarity,
and remote HII, as well as estimates of progenitor properties.


\begin{theacknowledgments}
   NP acknowledges partial support from STScI, through DDRF grant
   \#82367, and from INAF - Observatory of Rome to attend this
   conference.  KWW wishes to thank the Office of Naval Research (ONR)
   for the 6.1 funding supporting this research.
\end{theacknowledgments}



\begin{thebibliography}{9}

\bibitem[]{Chevalier82a} Chevalier, R.~A.~1982a, ApJ, 259, 302

\bibitem[]{Chevalier82b} Chevalier, R.~A.~1982b, ApJl, 259, L85

\bibitem[]{Chevalier98} Chevalier, R.~A.~1998, ApJ, 499, 810

\bibitem[]{Chu99} Chu, Y.-H., Caulet,  A., Montes, M.~J., Panagia, N., Van Dyk, S.~D., \& Weiler, K.~W.~1999, ApJl, 512, L51

\bibitem[]{Filippenko94} Filippenko, A., Matheson, T., \& Barth, A.~1994, AJ, 108, 222

\bibitem[]{Fransson98} Fransson, C., \& Bj\"orgsson, C.-I.~1998, ApJ, 509, 861

\bibitem[]{Immler01} Immler, S., Aschenbach, B., \& Wang, Q.~D.~2001, ApJl, 561, L107

\bibitem[]{Montes97} Montes, M.~J., Weiler, K.~W., \& Panagia, N.~1997, ApJ, 488, 792

\bibitem[]{Montes98}  Montes, M.~J., Van Dyk, S.~D., Weiler, K.~W., Sramek, R.~A., \& Panagia, N.~1998, ApJ, 506, 874

\bibitem[]{Montes00} Montes, M.~J., Weiler, K.~W., Van Dyk, S.~D., Sramek, R.~A., Panagia, N., \& Park, R.~2000, ApJ, 532, 1124

\bibitem[]{Natta84} Natta, A., \& Panagia, N.~1984, ApJ, 287, 228

\bibitem[]{Nomoto94}  Nomoto, K., Yamaoka, H., Pols, O.R., van den Heuvel, E., Iwamoto, K., Kumagai, S., Shigeyama, T.~1994, Nature, 371, 227

\bibitem[]{Panagia2006} Panagia, N., Van Dyk, S.~D., Weiler, K.~W., Sramek, R.~A.,  Stockdale, C.~J., \& Murata, K.~J.~2006, ApJ, 646, 369 

\bibitem[]{Podsiadlowski92} Podsiadlowski, Ph., Joss, P.~C., \& Hsu, J.~J.~L.~1992, ApJ, 391, 246

\bibitem[]{Podsiadlowski93} Podsiadlowski, Ph., Hsu, J., Joss, P., \& Ross, R. 1993, Nature, 364, 509

\bibitem[]{Ryder93} Ryder, S.~D., Staveley-Smith, L., Dopita, M., Petre, R., Colbert, E., Malin, D., \& Schlegel, E. 1993, ApJ, 417, 167

\bibitem[]{Ryder04} Ryder, S.~D., Sadler, E.~M., Subrahmanyan, R., Weiler, K.~W., Panagia, N., \& Stockdale, C.~J.~2004, MNRAS, 349, 1093

\bibitem[]{Schwarz96} Schwarz, D.~H., \& Pringle, J.~E.~1996, MNRAS, 282, 1018

\bibitem[]{Spyromilio94} Spyromilio, J.~1994, MNRAS, 266, 61

\bibitem[]{Sramek84} Sramek, R.~A., Panagia, N., \& Weiler, K.~W.~1984, ApJl, 285, L59

\bibitem[]{Sramek03} Sramek, R.~A., \& Weiler, K.~W.~2003, Supernovae and Gamma-Ray Bursters, ed.~K.~Weiler (Berlin: Springer-Verlag LNP 598) p.~145

\bibitem[]{VanDyk93a} Van Dyk, S.~D., Sramek, R.~A., Weiler, K.~W., \& Panagia, N. 1993a, ApJ, 409, 162

\bibitem[]{VanDyk93b} Van Dyk, S.~D., Sramek, R.~A., Weiler, K.~W., \& Panagia, N.~1993b, ApJl, 419, L69

\bibitem[]{VanDyk94}  Van Dyk, S.~D., Weiler, K.~W., Sramek, R., Rupen, M., \& Panagia, N.~1994, ApJl, 432, L115

\bibitem[]{Weiler86} Weiler, K., Sramek, R., Panagia, N., van der Hulst, J., \& Salvati, M.~1986, ApJ, 301, 790

\bibitem[]{Weiler90} Weiler, K.~W., Panagia, N., \& Sramek, R.~A.~1990, ApJ, 364, 611

\bibitem[]{Weiler91} Weiler, K., Van Dyk, S., Panagia, N., Sramek, R., \& Discenna, J.~1991, ApJ, 380, 161

\bibitem[]{Weiler92a} Weiler, K., Van Dyk, S., Pringle, J., \& Panagia, N.~1992a, ApJ, 399, 672

\bibitem[]{Weiler92b}  Weiler, K., Van Dyk, S., Panagia, N., \& Sramek, R.~1992b, ApJ, 398, 248


\bibitem[]{Weiler01} Weiler, K.~W., Panagia, N., \& Montes, M.~J.~2001, ApJ, 562, 670 

\bibitem[]{Weiler02} Weiler, K.~W., Panagia, N., Montes, M.~J., \& Sramek, R.~A.~2002, Ann. Rev. A \& Ap, 40, 387

\bibitem[]{Williams02} Williams, C.~L., Panagia, N., Lacey, C.~K., Weiler, K.~W., Sramek, R.~A., \& Van Dyk, S.~D.~2002, ApJ, 581, 396


\end{thebibliography}
\end{document}